# Dimension Effect of Nanocarbon Precursors on Diamond Synthesis and Transformation Mechanism under Extreme Conditions

Jiaxin Ming,[1,2] Jingyi Tian,[3] Liming Zhao,[4] Jiayin Li,[1,2] Guoshuai Du,[1,5] Lixing Kang,[4] Zheng Hu,[3,*] and Yabin Chen[1,2,5,*]

[1]*Advanced Research Institute of Multidisciplinary Sciences (ARIMS), Beijing Institute of Technology, Beijing 100081, China*

[2]*School of Chemistry and Chemical Engineering, Beijing Institute of Technology, Beijing 100081, China*

[3]*School of Chemistry and Chemical Engineering, Nanjing University, Nanjing 210023, China*

[4]*Key Laboratory of Multifunctional Nanomaterials and Smart Systems, Advanced Materials Division, Suzhou Institute of Nano-Tech and Nano-Bionics, Chinese Academy of Sciences, Suzhou 215123, China*

[5]*School of Aerospace Engineering, Beijing Institute of Technology, Beijing 100081, China*

[*]Correspondence and requests for materials should be addressed to: zhenghu@nju.edu.cn (Z.H.); chyb0422@bit.edu.cn (Y.C.).

**Abstract**

Diamond holds significant promise for a wide range of applications due to its exceptional physicochemical properties. Investigating the controlled diamond preparation from nanocarbon precursors with varying dimensions is crucial to optimize the transition conditions and even elucidate the daunting transformation mechanism, however, this remains outstanding challenge despite considerable effort. Herein, we report the imperative dimension effect of nanocarbon precursors on diamond synthesis and physical mechanism under high temperature and high pressure, by comparing the distinct transition processes of zero-dimensional (0D) carbon nanocages (CNCs) and one-dimensional (1D) carbon nanotubes (CNTs) from conventional graphite. The optical and structural characterizations evidently demonstrated that both 0D CNCs and 1D CNTs first undergo collapse and graphitization, followed by the formation of mixed amorphous carbon with embedded diamond clusters, eventually leading to cubic diamond. The plotted pressure-temperature diagram exhibits the unique dimension effect of carbon nanomaterials to diamond transformation. These results provide valuable insights into the phase transition mechanisms of diamond synthesis and its derivatives under extreme conditions.

**Introduction**

Cubic diamond (CD), with uniquely full $sp^3$ C-C bond, possesses exceptional physical properties, including its optical color center and ultrawide bandgap, elastic modulus (~1 TPa), ultrahigh thermal conductivity (2200 W m$^{-1}$K$^{-1}$ at 300 K), and biocompatibility.[1-4] Many CD-based derivatives have emerged these years with superior properties to the intrinsic CD phase, including polycrystalline diamond (pc-D),[5,6] nanotwinned diamond (nt-D),[7] and paracrystalline diamond (p-D).[8] For instance, it has been reported that the synthesized pc-D and nt-D surpass single-crystal CD in both mechanical hardness and toughness, owing to coherently interfaced polytypes.[9] Specifically, p-D, with medium range-order structure, presents the comparable hardness with single-crystal CD.[8] Therefore, the controllable synthesis of diamond and understanding its transformation mechanism have long been key focuses in the field of carbon materials.[10]

Numerous carbon precursors with different dimensions have been developed to prepare diamond and its derivatives with varying properties under specific conditions.[11] In 1955, Bundy F. P. et al. firstly achieved man-made diamond from three-dimensional (3D) graphite under high pressure and high temperature (HPHT).[12] Importantly, the empirical phase diagram of graphite has been plotted based on the systematical analysis of optimal pressure and temperature conditions.[13] It was found that zero-dimensional (0D) fullerene ($C_{60}$ and $C_{70}$)[14,15] and carbon onions[16] can transform to the functional diamond-like materials, such as the novel p-D from $C_{60}$ under 30 GPa and 1600 K.[8] Moreover, one-dimensional (1D) carbon nanotubes (CNTs) and nanofibers (CNFs) have been successfully used to prepare the CD phase,[17] and 1D CNFs undergo structural transition under 17 GPa and 2000 K, confirmed by the Raman and transmission electron microscopy (TEM) results.[18] Two-dimensional (2D) graphene can form ultrathin diamond, namely diamane, under high pressure, evidenced by its sizable bandgap and abrupt resistance.[19] Despite these advancements, phase diagram of diamond synthesized from various carbon allotropes remains largely unexplored.

Furthermore, the transition pathway and mechanism are crucial for the controlled preparation of diamond under extreme conditions. So far, plentiful models have been proposed to explain the graphite-diamond transformation, including coherent interfaces,[20] nucleation mechanism,[21,22] concerted transformation,[23,24] wave-like buckling and slipping mechanism.[25] Several studies found that 1D CNTs tend to graphitize first under high pressure, while further increased temperature can lead to the formation of diamond structures[26,27]. Additionally, it is proposed that for single-walled CNTs with small diameter distribution, bond formation and polymerization occur before graphitization under HPHT.[17] To the best of our knowledge, carbon nanocages (CNCs),[28] as the 0D paradigm of carbon nanomaterials, have never been used for diamond synthesis. The microscopic mechanism and physical properties of the formed diamonds from 0D CNCs and 1D CNTs keep largely unexplored.

In this study, we systematically studied the transition processes of various nanocarbon precursors with specific dimensions, including 0D CNCs and 1D CNTs under extreme conditions realized by our established laser-heating diamond anvil cell (LHDAC) system. Unlike 2D layered graphite, optical and TEM results revealed that 0D CNCs and 1D CNTs display the distinct transition pathway: from collapsing, graphite-like (GL) carbon, amorphous phase to eventual CD phase. The detailed pressure and temperature (*P-T*) diagram of 0D CNCs and 1D CNTs was further mapped out. We believe these results should shed light on the controlled preparation and structural modulation of diamond under HPHT.

**Results and discussion**

1D CNTs precursor features a dense network structure composed of randomly oriented nanotubes,[29] and its average diameter is around 7 nm as shown in Figures 1a and S1a-b. In comparison, 0D CNCs precursor exhibits uniquely hollow hierarchical structure with an interior cavity size of ~25 nm,[30] and spontaneously aggregates into a loose powder due to van der Waals coupling in Figures 1d and S1d-e. Figure 1b and 1e present optical micrographs of CNTs and CNCs samples loaded in DAC after laser heating under high pressure. Both precursors, CNTs at 17.3 GPa and CNCs at 30.4 GPa, appear opaque under high pressure and room temperature. Once shone by the high-power laser beam, either CNTs or CNCs suddenly became transparent, and the heating temperatures were calibrated as about 1520 and 1550 K, respectively, indicating the potential phase transition accompanied with diamond formation.

Importantly, the laser-heating approach allowed us to selectively regulate the elevated temperature (i.e. laser power) and thus the diamond pattern together with the desired phase boundaries between unconverted carbon and diamond, facilitating the investigations into transformation mechanism (we will discuss this later). In this manner, the nominal diamond "dot", "wire" and "film" can be reliably produced by manually controlling the trajectory of laser motion, as displayed in Figures S2-S3. The extracted spectral width of Raman $T_{2g}$ peak can be as narrow as 4.7 cm$^{-1}$ (Figure S2). In contrast, without being subjected to high temperature, 0D CNCs or 1D CNTs precursors were reversibly preserved once totally decompressed.

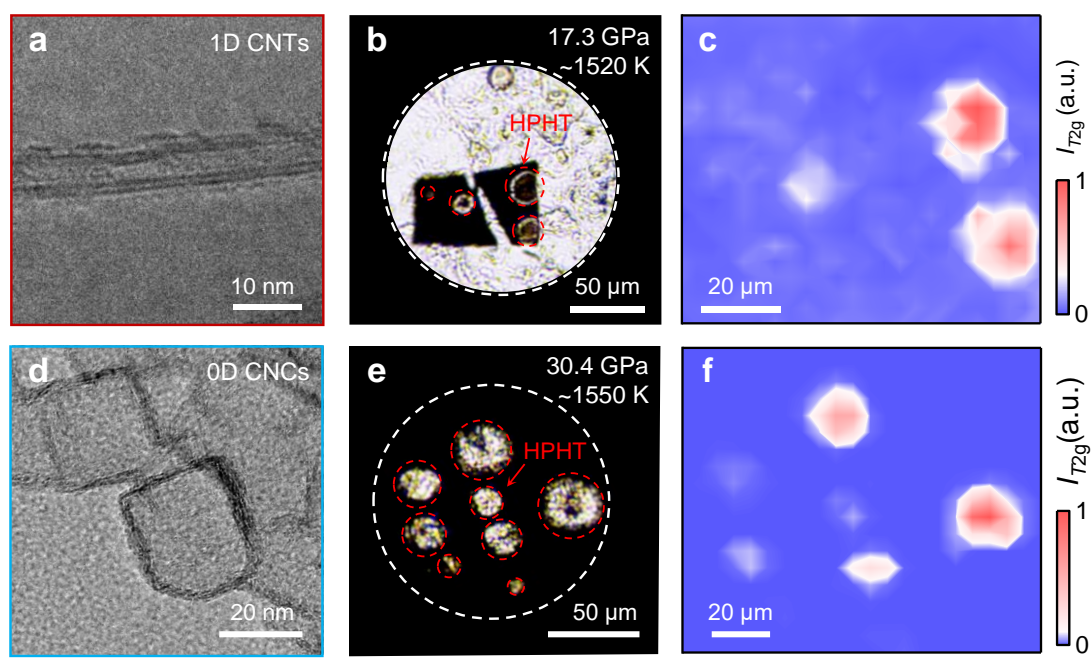

**Figure 1. Diamond synthesis from 1D CNTs and 0D CNCs precursors under HPHT.** (**a**) Representative TEM image of the pristine 1D CNTs. (**b**) Optical image of 1D CNTs in DAC after heating under high pressure. (**c**) Intensity mapping of Raman $T_{2g}$ peak for CNTs recovered from HPHT. (**d**) Representative TEM image of 0D CNCs. (**e**) Optical image of CNCs after 30.4 GPa and ~1550 K. (**f**) Intensity mapping of $T_{2g}$ peak for CNCs after HPHT.

Intriguingly, it was found that the formed transparent phase can be quenched to ambient conditions, and present the distinguished optical properties. First, the obtained Raman spectrum showed a strong and sharp peak at 1332 cm$^{-1}$, corresponding to the first-order phonon vibration ($T_{2g}$) of CD.[31] The Raman mapping results in Figure 1c and 1f are reasonably consistent with the laser-heated patterns. Second, photoluminescence (PL) mapping results of the same recovered samples revealed a strong peak at 738 nm, resulting from the characteristic SiV$^-$ center within the diamond regions (Figures S4). The silicon dopants possibly originated from the synthesis process of carbon precursors.

Optical properties of the quenched samples under various pressure and temperature conditions were investigated, including Raman, PL, and absorption spectra. Figure 2a and 2b display the representative Raman spectra of recovered samples from 1D CNTs and 0D CNCs after low (LT), moderate (MT) and high temperature treatments under high pressure, respectively. Interestingly, Raman results of three quenched products from each CNTs and CNCs are very similar, classified into three distinct phases as GL structure, amorphous carbon (AC) and CD. For the GL structure quenched from HPLT condition, Raman spectrum usually exhibits two strong peaks: G band ($E_{2g}$ mode) from the stretching vibration of C-C bonds and D band associated with the inevitable defects or disorder. According to the peak area ratio of D to G band ($S_D/S_G$), the size $L_a$ of graphitic cluster can be estimated from the empirical formula $L_a(\text{nm})=A\lambda_{\text{laser}}^4(S_D/S_G)^{-1}$,[32] where $\lambda_{\text{laser}}$ is laser wavelength (405 nm in our study), $A\sim2.4\times10^{-10}$ nm$^{-3}$ is a constant. The calculated cluster size ~5.0 nm synthesized by CNTs (20.4 GPa, 1350 K) is larger than ~3.0 nm from CNCs (17.0 GPa, 1050 K) in Figure S5. As $P$ and $T$ increase further, AC phase formed, and its Raman spectrum along with a strong fluorescence background mainly features four peaks: the characteristic T (~1080 cm$^{-1}$) and A (~1530 cm$^{-1}$) bands of amorphous $sp^3$ carbon,[33,34] D (~1404 cm$^{-1}$) and G (~1600 cm$^{-1}$) bands from $sp^2$ carbon,[35] as shown in Figure S6. When the critical conditions are approached, the CD phase is achieved eventually, and the exclusive Raman peak at 1332 cm$^{-1}$ belongs to its first-order $T_{2g}$ mode. The fitted full-width at half maximum of the synthesized diamond from CNTs (7.4 cm$^{-1}$) and CNCs (5.4 cm$^{-1}$) are comparable with natural gemstone, indicating the high crystalline quality of our synthesized diamonds. The vanished G and D bands from CNTs or CNCs precursors evidently proved the great transformation efficiency to the diamond phase.

In Figure 2c, G-band frequency of 1D CNTs increases linearly up to 13 GPa, slightly higher than 8 GPa observed for 0D CNCs. The extracted pressure coefficient of 3.8 cm$^{-1}$/GPa for CNTs is consistent with MWCNTFs result (3.6 cm$^{-1}$/GPa),[18] while relatively smaller than 4.3 cm$^{-1}$/GPa for our CNCs and graphite (4.0 cm$^{-1}$/GPa) in literature.[36] When pressure exceeds ~13 GPa, the slope for CNTs begins to level off, potentially corresponding to their structural collapse along the radial direction. In contrast, the slope of CNCs changes at a lower pressure around 8 GPa, attributed to the weaker compression resistance of the large internal hollow structure of CNCs (Figure 1d). After releasing the pressure, Raman spectrum of the original samples came back, indicating that the structural changes under high pressure alone are reversible (Figures S7-S8). Figure 2d shows the measured absorption spectra of diamonds synthesized from CNTs and CNCs. The indirect optical bandgaps, derived by Tauc plot approach [$(\alpha h\nu)^{1/2}$ vs. $h\nu$],[37] are determined as 1.36 and 1.24 eV for CNTs and CNCs precursors respectively, dramatically smaller than that of natural diamond (~5.47 eV).[38]

Furthermore, PL properties of our synthesized diamond from two precursors were explored, as shown in Figure 2e-f. PL spectrum, excited by a 405 nm laser at 300 K, shows two features at 575 and 738 nm, corresponding to the zero phonon lines (ZPL) of NV$^0$ and SiV$^-$ color centers, respectively. The additional peak at 637 nm only appears under 532 nm excitation, attributed to ZPL of NV$^-$ center. It is reported that NV$^-$ center becomes a dark state and transited to NV$^0$ under ultraviolet excitation,[39] which occurred in our case using 405 nm laser. Importantly, the PL range of 400-550 nm becomes more profound at 100 K, accompanied with many emergent sharp peaks (Figure S8). These features at 468, 495, and 505 nm are quantitatively consistent with the ZPLs of TR12, H4 (4N+2V), and H3 defect centers,[40] suggesting that Si and N are the dominant dopants in our prepared diamonds. This result was confirmed by energy-dispersive x-ray spectrum as well (Figure S10). Moreover, ZPL peak of SiV$^-$ center becomes weakened, broadened and redshifted as temperature increasing in Figure 2f and S9, primarily due to lattice expansion and electron-phonon coupling.[41] The ZPL energy can be well fitted with the modified Varshni model $E(T)=E_0-\alpha T^4/(T+\beta)^2$, where $E_0$ is the intrinsic energy at absolute zero temperature, and $\alpha$ and $\beta$ are two fitting parameters.[42] The extracted values of $E_0$ = 1.682 eV, $\alpha$ = 4.23 × 10$^{-8}$ eV/K$^2$, and $\beta$ = 11.44 K are well consistent with literature results.[42]

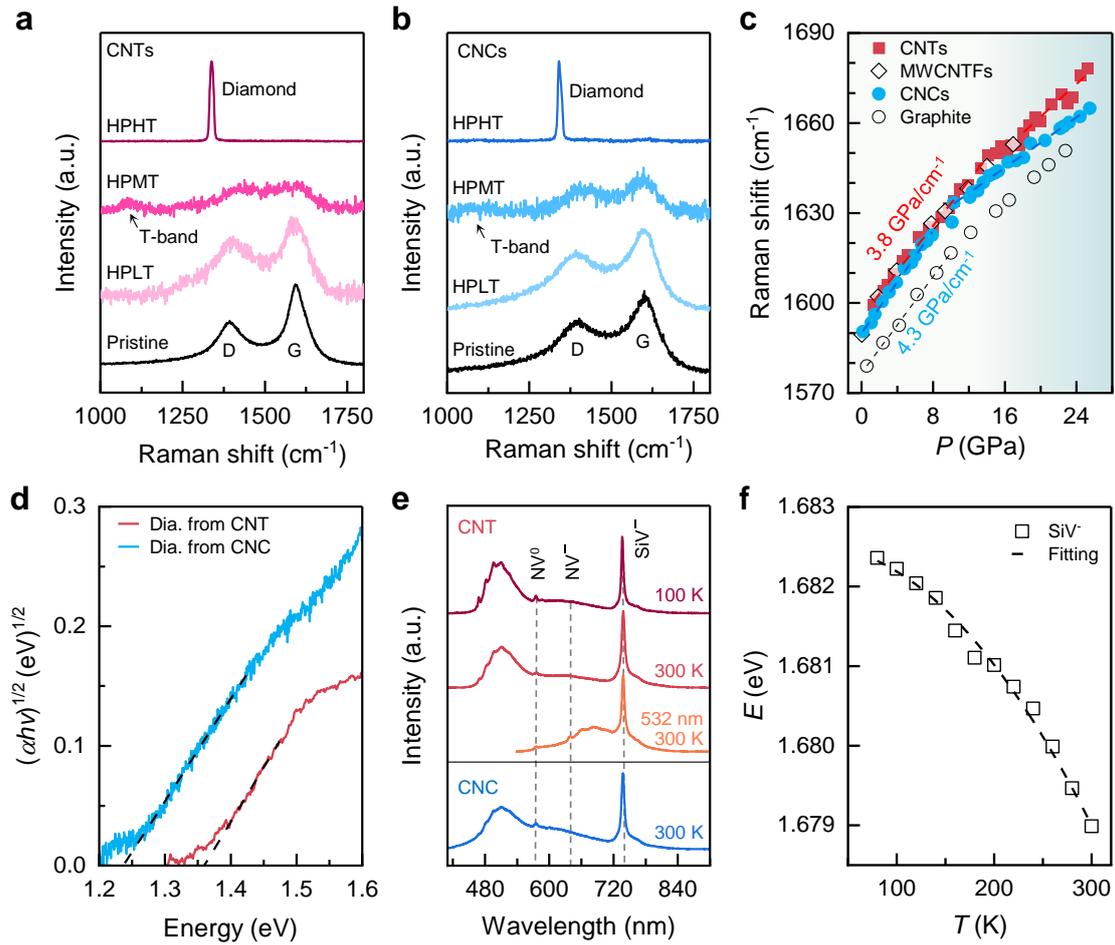

**Figure 2. Optical properties of the synthesized samples by 1D CNTs and 0D CNCs.** (**a**) Raman spectra of the pristine precursor and recovered samples synthesized by 1D CNTs. HPLT (20.4 GPa/1350 K), HPMT (20.2 GPa/1500 K) and HPHT (17.4 GPa/1650 K) conditions for each CNTs precursor are used to prepare GL carbon, AC, and CD samples, respectively. (**b**) Raman spectra of the pristine precursor and recovered samples from 0D CNCs. HPLT, HPMT and HPHT conditions for CNCs are 17 GPa/1050 K, 22.7 GPa/1400 K, and 20.7 GPa/1500 K, respectively. (**c**) Pressure-dependence of G band frequency of CNCs (blue), CNTs (red), graphite (black circle), and multi-walled carbon nanotubes fibers (MWCNFs, black diamond).[18] The dashed lines mean the linear fitting results within a certain pressure range. (**d**) Absorption spectra of diamonds synthesized by 1D CNTs (red) and 0D CNCs (blue), and the extracted band gap 1.36 and 1.24 eV by Tauc plot method, respectively. (**e**) PL spectra of diamond synthesized by 1D CNTs and 0D CNCs. The laser wavelength of excitation source is 405 nm, expect one curve by 532 nm. The color centers of $NV^0$, $NV^-$ and $SiV^-$ are marked with dashed lines. (**f**) Temperature-dependent ZPL energy of $SiV^-$ color center using 405 nm laser. The dashed line represents the fitting result.

To explore their structural transition mechanism, atomic nanostructures of the recovered samples from various conditions were further examined using high-resolution TEM (HRTEM) and selected area electron diffraction (SAED). In Figure 3a-b, the GL and deformed structure were observed around the thermal-diffusion region. The measured interlayer distance of 3.57 Å closes to the characteristic $d_{(002)}$ = 3.4 Å of crystalline graphite, proved through the fast Fourier transform (FFT) pattern in Figure 3b. The curved and opaque product should be raised by the collapsed nanotubes, followed by the fracture at edges. This assumption can be well verified with the prominent G and D bands of Raman spectrum in Figure 2a. As $T$ increases, a mixture of AC and nano-CD phases was quenched to ambient conditions. The FFT pattern of AC is totally diffusive without any ordering feature in Figure 3d, in good agreement with literature.[43] In comparison, the nano-CD phase exhibited the remarkable short-range order along (111) orientation, and the bright spots of FFT pattern originated from its (111) plane with the measured $d_{(111)}$= 2.09 Å, although this phase can't show any Raman signal in Figure 2a. The formation of AC phase is presumably due to the enhanced atomic disorder in the GL structure with rising temperature, which also facilitates the nucleation of diamonds. Figure 3f displays the typical HRTEM result of the recovered CD structure under 17.4 GPa and 1650 K, confirmed by the sharp patterns in SAED result. Figure 3g shows the obtained FFT image of crystalline CD phase along [$01\bar{1}$] zone axis. Three conspicuous spots with $d$-spacings of 2.02, 1.71 and 2.08 Å correspond to ($1\bar{1}\bar{1}$), (200) and (111) planes of CD, respectively. Moreover, the angles of ($1\bar{1}\bar{1}$)/(200) and (111)/(200) planes are measured to be 52.3 ° and 53.3°, respectively, quantitatively closed to the calculated values of CD (PDF#06-0675). These results indicate that CD phase forms through the further growth of nano-diamonds under HPHT. Basically, we propose that the transition of CNTs to diamond first passes through a GL phase, then progresses to AC matrix with nanodiamond clusters, ultimately stabilizing into a CD structure (Figures 3 and S11). The obtained size distribution of CD product falls in 100–250 nm via TEM analysis (Figure S13a).

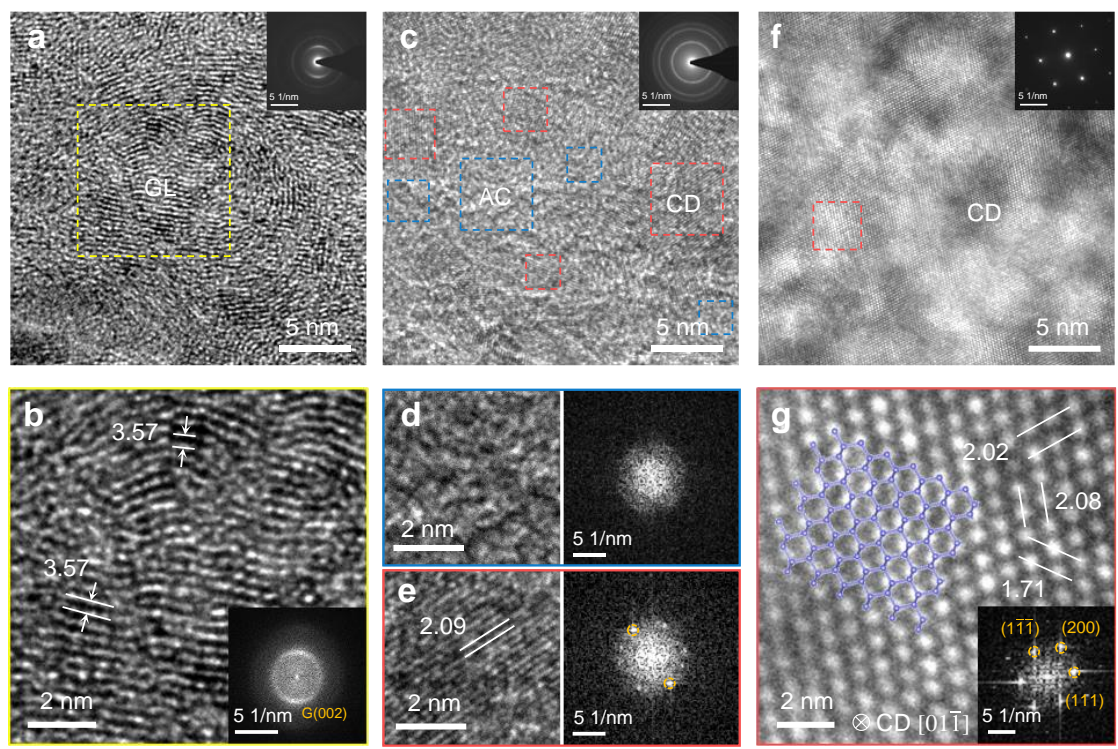

**Figure 3. HRTEM results of the quenched samples from 1D CNTs under different conditions.** (**a**) TEM image of GL structure synthesized at 17.4 GPa and low temperature. Inset is the corresponding SAED pattern. (**b**) HRTEM image of the yellow square marked in **a**. The corresponding FFT pattern presents a diffusive ring around 3.57 Å, suggesting a deformed and layered GL structure. (**c**) TEM and SAED images of the obtained mixture of AC and CD clusters after 20.2 GPa and 1500 K. (**d**) TEM and FFT results of AC phase obtained from the blue area in **c**. (**e**) TEM and FFT results of nano-polycrystalline CD from the red area in **c**. The measured 2.09 Å means (111) plane of CD. (**f**) TEM and SAED results of high-quality CD structure after 17.4 GPa and 1650 K. (**g**) HRTEM image and FFT result of CD phase from the red square in **f**. The labeled 2.02, 1.71 and 2.08 Å correspond to (1$\bar{1}\bar{1}$), (200) and (111) planes of CD, respectively.

HRTEM and SAED characterizations of the recovered samples from 0D CNCs under HPHT has been performed as well, as shown in Figure 4. Similar to 1D CNTs, we found that CNCs first transform into GL structure with *d*-spacing of 3.71 Å at relatively low temperatures (Figure 4a). However, these GL carbon phase appears more oriented and linear, without any apparent curves shown in that synthesized by CNTs (Figure 3a-b). This phenomenon can be reasonably understood, because the unique inner cavity of CNCs is much larger than the diameter of CNTs precursors (Figure 1). Subsequently, two mixed phases of AC and nano-CD clusters closely resemble that of CNTs after HPMT (Figure 3c). The measured distance ~2.02 Å can be assigned as the interplanar spacing of (111) planes in Figure 4e. With further increase of temperature, AC regions gradually vanished, and the CD domains grew and eventually formed a large crystalline size, as shown in Figure 4f-g. HRTEM image along [01$\bar{1}$] zone axis shows that the intrinsic cubic symmetry of diamond crystals. The extracted distances of 1.73, 2.05 and 2.10 Å correspond to the (200), (1$\bar{1}\bar{1}$) and (111) planes of CD lattice, respectively. Therefore, the phase transition process of CNCs is similar to that of CNTs (Figures 4 and S12). The mean size of the obtained diamonds from 0D CNCs lies within 90–200 nm, slightly smaller than those synthesized from 1D CNTs (Figure S13b). In addition, we observed twin boundaries and diamond polytypes in the synthesized diamond, as shown in Figure S14.

Next, the *P-T* diagram of 0D CNCs and 1D CNTs can be tentatively mapped out by classifying the synthesized samples under HPHT based on their optical and TEM results, as shown in Figure 5a. Apparently, both CNTs and CNCs exhibit three distinct phase structures under extreme conditions. For 1D CNTs, the phase diagram can be divided into three regions: (1) the region of *sp$^2$*-bonded GL structure synthesized at HPLT (<1450 K); (2) the region of *sp$^2$*- and *sp$^3$*-bonded AC phase in coexistence with *sp$^3$*-bonded nanodiamond clusters recovered at HPMT (1450-1550 K); (3) the region of *sp$^3$*-bonded CD phase synthesized under HPHT (>1550 K). In comparison, 0D CNCs also undergo a similar transformation pathway: the GL phase synthesized at HPLT (<1300 K), followed by the metastable AC phase and nanodiamond at HPMT (1300-1500 K), and CD as the temperature rises further (>1500 K). Furthermore, high pressure (like 35.5 GP at 300 K) or high temperature alone (below 10 GPa, even at temperatures exceeding 3000 K) is insufficient to quench the diamond products to ambient condition. According to the obtained phase diagram, it can be seen that the *P* and *T* conditions required for CNCs to synthesize diamonds are slightly milder than those for CNTs below ~20 GPa, partly because of its larger cavity and more disordered structures

(Figure S5).

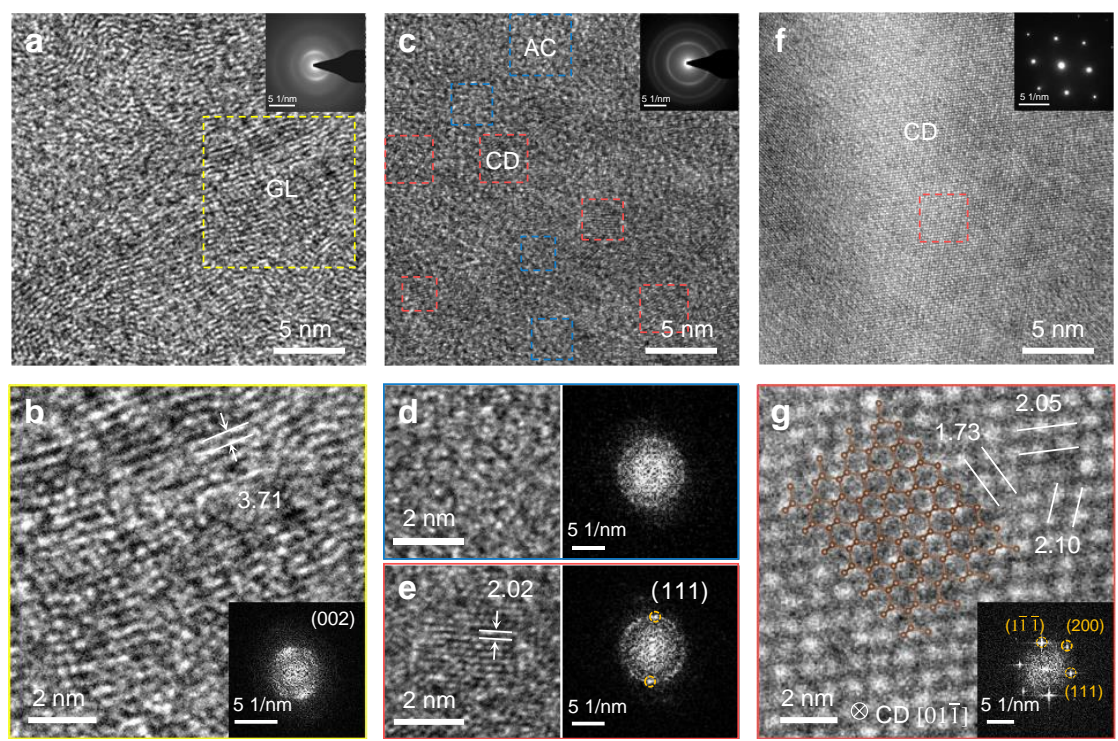

**Figure 4. HRTEM results of the quenched samples from 0D CNCs under different conditions.** (**a**) TEM image of GL structure after 22.7 GPa and low temperature. Inset is the corresponding SAED pattern. (**b**) HRTEM result of the yellow square marked in **a**. The corresponding FFT pattern presents a diffusive ring around 3.71 Å, suggesting a layered graphite structure. (**c**) TEM and SAED images of a mixture of AC and diamond clusters after 22.7 GPa and 1400 K. (**d**) TEM and FFT results of AC phase obtained from the blue area in **c**. (**e**) TEM and FFT results of nanocrystalline CD phase from the red area in **c**. The measured 2.02 Å means (111) plane of CD. (**f**) TEM and SAED results of the crystalline CD obtained after 20.7 GPa and 1500 K. (**g**) HRTEM and FFT result of CD phase from the red square in **f**.

Figure 5b illustrates the transition diagram of 1D CNTs and 0D CNCs under HPHT, compared with other carbon allotropes such as 0D $C_{60}$ and 2D graphite. The $P$ and $T$ conditions for diamond synthesis from CNTs and CNCs are generally similar to those of 0D $C_{60}$, while significantly lower than those required for 2D layered graphite. More severe conditions are needed to trigger 2D graphite-CD transformation, attributed to its intrinsic absence of $sp^3$ configuration unlike other carbon nanomaterials. Additionally, the transition pathways for CNTs and CNCs remarkably differ from that of 2D graphite, proving the effect of precursor dimensions. Sequentially, the hollow 0D CNCs and 1D CNTs can collapse and then undergo graphitization under high pressure, densely packed $sp^3$-bonded AC with embedded diamond appear once temperature effect plays the key role, the atomic arrangement becomes more ordered as temperature increases further, and ultimately the CD phase forms.

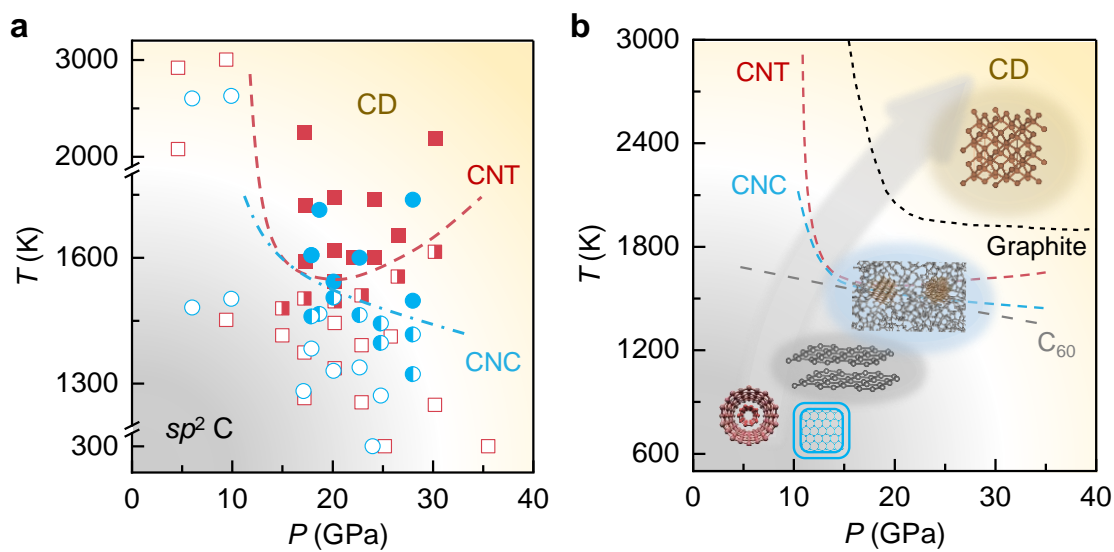

**Figure 5. Structural transformation from 0D CNCs and 1D CNTs precursors to diamond under HPHT.** (**a**) *P-T* phase diagram of 0D CNCs (circles) and 1D CNTs (squares). Opened, dual-coloured and filled symbols denote the GL, AC mixed with diamond structure and CD, respectively. (**b**) Schematic diagram of phase transitions of 0D CNCs and 1D CNTs under HPHT, compared with 0D $C_{60}$ and 2D graphite. The red, blue, gray and black dashed lines represent the phase boundaries between CNTs, CNCs, $C_{60}$,[14] graphite[13] and the stable CD, respectively.

**Conclusion**

In summary, we investigated the structural phase transitions of 1D CNTs and 0D CNCs with different dimensions under extreme conditions. By comparatively analyzing the plotted *P-T* diagrams of CNTs and CNCs together with 2D graphite and 0D $C_{60}$, we found that the consistent transition pathways of CNTs and CNCs are obviously different from that of graphite due to dimension effect. 1D CNTs and 0D CNCs initially undergo collapse under high pressure and then form a GL phase when temperature begins to play a role, which is followed by the AC phase with embedded diamond nanoclusters and finally CD phase as temperature increases. This study may lay a solid foundation for the HPHT synthesis of carbon nanomaterials and transition mechanisms of diamond.

**EXPERIMENTAL METHODS**

**Synthesis of carbon precursors**

The controlled preparation of CNTs involved a continuous floating catalyst chemical vapor deposition (CVD) and followed by a densification procedure. The optimized procedure can be found elsewhere.[29] In brief, a precursor solution of ethanol, 1.2 vol% ferrocene, and 0.4 vol% thiophene was injected into a hot furnace at 1100 °C under a mixed Ar/$H_2$ atmosphere. Thermal decomposition of the feedstock facilitated CNT growth, forming an aerogel that floated with the gas flow and was continuously collected on a rotating roller as a CNTs sponge. A large-sized CNTs film was eventually obtained by spraying ethanol into CNTs network for generating capillary forces that reduced its thickness.

The CNCs were synthesized by using *in-situ* MgO template method, with benzene as the precursor in a tubular furnace, similar to our previous report.[30] Basic magnesium carbonate ($4MgCO_3·Mg(OH)_2·5H_2O$) was heated to 800 °C at a rate of 10 °C $min^{-1}$ in argon atmosphere, and benzene (1 mL) was injected over a period of 10 min via a constant-flow pump. Following 1 h of heating at 800 °C and cooling down to room temperature, the product was then treated with $H_2SO_4$ aqueous solution to remove MgO template, resulting in the desired hollow nanocages.

**High pressure and high temperature experiments**

The high pressure was generated via a symmetric DAC with 400 μm culet size. The sample chamber was prepared by laser drilling a ~150 μm hole into a T301 steel gasket, which had a pre-indented thickness of ~40 μm. The solid sodium chloride was utilized as the pressure transmitting medium, and layered mica as thermal insulator. Pressure was calibrated using the shift of R1 fluorescence peak of a ruby ball, placed near the carbon precursors in the sample chamber. The CNT film was shaped into approximately 70×70 mm square, and the CNC powder was compressed into pellets using a tablet press. For the high pressure and high temperature synthesis of diamond phase, a homemade double-sided laser heating system, equipped with 1070 nm lasers ($P_{max}$ = 100 W) on each side, was used to heat the carbon precursors. The heating temperature was calibrated by fitting the acquired blackbody radiation curve (530-800 nm normally) of the sample with Planck's law (Figure. S15).

**Optical spectroscopy and structural measurements**

All Raman and PL spectra were collected in backscattering configuration using a Horiba iHR500 spectrometer equipped with the advanced CCD camera. The aligned 405 or 532 nm laser as excitation source illuminated the carbon precursors in DAC through a super long-working distance objective (SLMPLN 50×, Olympus). The optical bandgap of the recovered samples was measured by fitting the absorption spectra through Tauc plot method. The homemade absorption system consists of a halogen lamp as light source (wavelength range of 360 to 2000 nm) and a grating spectrometer (HRS-300SS, Princeton Instruments). X-ray photoelectron spectroscopy measurement was conducted on an Axis Supra+ system with an Al Kα X-ray source ($hv$ = 1486.8 eV). HRTEM and the related SAED results were obtained with a FEI Titan Cube apparatus at an acceleration voltage of 200 kV.

**ASSOCIATED CONTENTS**

**Supporting Information**

The Supporting Information is available free of charge online.

Optical characterization of the original samples; Optical micrographs and Raman mapping of high-temperature and high-pressure products; Pressure dependence of Raman spectra for the pristine samples; Temperature dependence of PL for diamond products; EDS images of synthesized diamond; TEM and SAED images of recovered samples; Grain size distribution of synthesized diamond; Temperature calibration using blackbody radiation.

**Author Contributions**

Y.C. conceived this research project and designed the experiments. J.T. and J.M. prepared CNCs advised by Z.H. L.Z. and J.M. prepared CNTs with the advice of L.K. J.M. and J.L. developed the laser heating system. J.M. and G.D. designed and established the optical setups. J.M performed all high pressure and high temperature experiments. J.M. processed the experimental data under the advice of Y.C. J.M. and Y.C. wrote the manuscript with the essential feedback of all co-authors. All authors have given approval to the final manuscript.

**Notes**

The authors declare no competing financial interest.

**Acknowledgements**

This work was financially supported by the National Natural Science Foundation of China (grant numbers 52072032, 52472040 and 12090031), and the 173 JCJQ program (grant No. 2021-JCJQ-JJ-0159).**Data Availability**

All data related to this study are available from the corresponding author upon reasonable request.

# REFERENCES

1  Lodahl, P., Mahmoodian, S. & Stobbe, S. Interfacing single photons and single quantum dots with photonic nanostructures. *Rev. Mod. Phys.* **87**, 347-400, (2015).

2  Carbery, W. P., Farfan, C. A., Ulbricht, R. & Turner, D. B. The phonon-modulated Jahn–Teller distortion of the nitrogen vacancy center in diamond. *Nat. Commun.* **15**, 8646, (2024).

3  Mochalin, V. N., Shenderova, O., Ho, D. & Gogotsi, Y. The properties and applications of nanodiamonds. *Nat. Nanotechnol.* **7**, 11-23, (2012).

4  Field, J. E. The properties of natural and synthetic diamond. *Academic, New York* (1992).

5  Tanigaki, K. *et al.* Observation of higher stiffness in nanopolycrystal diamond than monocrystal diamond. *Nat. Commun.* **4**, 2343, (2013).

6  Irifune, T., Kurio, A., Sakamoto, S., Inoue, T. & Sumiya, H. Ultrahard polycrystalline diamond from graphite. *Nature* **421**, 599-600, (2003).

7  Huang, Q. *et al.* Nanotwinned diamond with unprecedented hardness and stability. *Nature* **510**, 250-253, (2014).

8  Tang, H. *et al.* Synthesis of paracrystalline diamond. *Nature* **599**, 605-610, (2021).

9  Yue, Y. *et al.* Hierarchically structured diamond composite with exceptional toughness. *Nature* **582**, 370-374, (2020).

10 Zeng, Z. *et al.* Synthesis of quenchable amorphous diamond. *Nat. Commun.* **8**, 322, (2017).

11 Sundqvist, B. Carbon under pressure. *Phys. Rep.* **909**, 1-73, (2021).

12 Bundy, F. P., Hall, H. T., Strong, H. M. & Wentorfjun, R. H. Man-made diamonds. *Nature* **176**, 51-55, (1955).

13 Bundy, F. P. *et al.* The pressure-temperature phase and transformation diagram for carbon; updated through 1994. *Carbon* **34**, 141-153, (1996).

14 Shang, Y. *et al.* Ultrahard bulk amorphous carbon from collapsed fullerene. *Nature* **599**, 599-604, (2021).

15 Shang, Y. *et al.* Enhancement of short/medium-range order and thermal conductivity in ultrahard $sp^3$ amorphous carbon by $C_{70}$ precursor. *Nat. Commun.* **14**, 7860, (2023).

16 Banhart, F. & Ajayan, P. M. Carbon onions as nanoscopic pressure cells for diamond formation. *Nature* **382**, 433-435, (1996).

17  Merlen, A. *et al.* High pressure–high temperature synthesis of diamond from single-wall pristine and iodine doped carbon nanotube bundles. *Carbon* **47**, 1643-1651, (2009).

18  Yang, X. *et al.* Diamond-graphite nanocomposite synthesized from multi-walled carbon nanotubes fibers. *Carbon* **172**, 138-143, (2021).

19  Ke, F. *et al.* Synthesis of atomically thin hexagonal diamond with compression. *Nano Lett.* **20**, 5916-5921, (2020).

20  Luo, K. *et al.* Coherent interfaces govern direct transformation from graphite to diamond. *Nature* **607**, 486-491, (2022).

21  Khaliullin, R. Z., Eshet, H., Kühne, T. D., Behler, J. & Parrinello, M. Nucleation mechanism for the direct graphite-to-diamond phase transition. *Nat. Mater.* **10**, 693-697, (2011).

22  Britun, V. F., Kurdyumov, A. V. & Petrusha, I. A. Diffusionless nucleation of lonsdaleite and diamond in hexagonal graphite under static compression. *Powder Metall. Met. Ceram.* **43**, 87-93, (2004).

23  Fahy, S., Louie, S. G. & Cohen, M. L. Theoretical total-energy study of the transformation of graphite into hexagonal diamond. *Phys. Rev. B* **35**, 7623-7626, (1987).

24  Scandolo, S., Bernasconi, M., Chiarotti, G. L., Focher, P. & Tosatti, E. Pressure-induced transformation path of graphite to diamond. *Phys. Rev. Lett.* **74**, 4015-4018, (1995).

25  Xie, H., Yin, F., Yu, T., Wang, J.-T. & Liang, C. Mechanism for direct graphite-to-diamond phase transition. *Sci. Rep.* **4**, 5930, (2014).

26  Silva-Santos, S. D. *et al.* From high pressure radial collapse to graphene ribbon formation in triple-wall carbon nanotubes. *Carbon* **141**, 568-579, (2019).

27  Tang, D. S. *et al.* Behavior of carbon nanotubes under high pressure and high temperature. *J. Mater. Res.* **15**, 560-563, (2000).

28  Wu, Q., Yang, L., Wang, X. & Hu, Z. Carbon-based nanocages: A new platform for advanced energy storage and conversion. *Adv. Mater.* **32**, 1904177, (2020).

29  Wang, H. *et al.* Ultra-lightweight and highly adaptive all-carbon elastic conductors with stable electrical resistance. *Adv. Funct. Mater.* **27**, 1606220, (2017).

30  Xie, K. *et al.* Carbon nanocages as supercapacitor electrode materials. *Adv. Mater.* **24**, 347-352, (2012).

31  Knight, D. S. & White, W. B. Characterization of diamond films by Raman spectroscopy. *J. Mater. Res.* **4**, 385-393, (1989).

32  Pimenta, M. A. *et al.* Studying disorder in graphite-based systems by Raman spectroscopy. *Phys. Chem. Chem. Phys.* **9**, 1276-1290, (2007).

33  Ferrari, A. C. & Robertson, J. Resonant Raman spectroscopy of disordered, amorphous, and diamondlike carbon. *Phys. Rev. B* **64**, 075414, (2001).

34  Bogdanov, K. *et al.* Annealing-induced structural changes of carbon onions: High-resolution transmission electron microscopy and Raman studies. *Carbon* **73**, 78-86, (2014).

35  Ferrari, A. C. & Robertson, J. Interpretation of Raman spectra of disordered and amorphous carbon. *Phys. Rev. B* **61**, 14095-14107, (2000).

36  Hanfland, M., Beister, H. & Syassen, K. Graphite under pressure: Equation of state and first-order Raman modes. *Phys. Rev. B* **39**, 12598-12603, (1989).

37  Tauc, J., Grigorovici, R. & Vancu, A. Optical properties and electronic structure of amorphous germanium. *Phys. Status Solidi B* **15**, 627-637, (1966).

38  Wort, C. J. H. & Balmer, R. S. Diamond as an electronic material. *Mater. Today* **11**, 22-28, (2008).

39  Beha, K., Batalov, A., Manson, N. B., Bratschitsch, R. & Leitenstorfer, A. Optimum photoluminescence excitation and recharging cycle of single nitrogen-vacancy centers in ultrapure diamond. *Phys. Rev. Lett.* **109**, 097404, (2012).

40  Zaitsev, A. M. *Optical properties of diamond: a data handbook*. (Springer Science & Business Media, 2013).

41  Chen, X. D. *et al.* Temperature dependent energy level shifts of nitrogen-vacancy centers in diamond. *Appl. Phys. Lett.* **99**, 161903, (2011).

42  Li, C. C. *et al.* Temperature dependent energy gap shifts of single color center in diamond based on modified Varshni equation. *Diamond Relat. Mater.* **74**, 119-124, (2017).

43  Zhang, S. *et al.* Narrow-gap, semiconducting, superhard amorphous carbon with high toughness, derived from $C_{60}$ fullerene. *Cell Rep. Phys. Sci.* **2**, 100575, (2021).